\documentclass[trackchanges,twocolumn]{aastex7}  %linenumbers,

\usepackage{amsmath,amstext}
\usepackage[T1]{fontenc}
\usepackage{multirow}
\usepackage[utf8]{inputenc} 
\usepackage{float}
\usepackage{graphicx}

\usepackage{tabularx} 
\usepackage{booktabs}

\usepackage{threeparttable}

\shorttitle{Validation of TOI-2267 d}
\shortauthors{Greklek-McKeon et al.}

\begin{document}

\title{Validation of a Third Earth-sized Planet in the TOI-2267 Binary System}

\correspondingauthor{Michael Greklek-McKeon}
\email{mgreklek-mckeon@carnegiescience.edu}

\author[0000-0002-0371-1647]{Michael Greklek-McKeon}
\affiliation{Earth and Planets Laboratory, Carnegie Institution for Science, Washington, DC 20015, USA}
\affiliation{Division of Geological and Planetary Sciences, California Institute of Technology, Pasadena, CA 91125, USA}
\email{michael@caltech.edu}

\author[0000-0002-0672-9658]{Jonathan Gomez Barrientos}
\affiliation{Division of Geological and Planetary Sciences, California Institute of Technology, Pasadena, CA 91125, USA}
\email{jgomezba@caltech.edu}

\author[0000-0002-5375-4725]{Heather A. Knutson}
\affiliation{Division of Geological and Planetary Sciences, California Institute of Technology, Pasadena, CA 91125, USA}
\email{hknutso2@caltech.edu}

\author[0000-0002-9350-830X]{Sebasti\'an Z\'u\~niga-Fern\'andez}
\affiliation{Astrobiology Research Unit, Universit\'e de Li\`ege, All\'ee du 6 Ao\^ut 19C, B-4000 Li\`ege, Belgium}
\email{sgzuniga@uliege.be}

\author[0000-0003-1572-7707]{Francisco J. Pozuelos}
\affiliation{Instituto de Astrof\'isica de Andaluc\'ia (IAA-CSIC), Glorieta de la Astronom\'ia s/n, 18008 Granada, Spain}
\email{pozuelos@iaa.es}

\author[0000-0001-9518-9691]{Morgan Saidel}
\affiliation{Division of Geological and Planetary Sciences, California Institute of Technology, Pasadena, CA 91125, USA}
\email{msaidel@caltech.edu}

\author[0000-0002-1422-4430]{W. Garrett Levine}
\affiliation{Department of Astronomy, Yale University, New Haven, CT 06511, USA}
\affiliation{Department of Earth, Planetary, and Space Sciences, University of California, Los Angeles, CA 90095, USA}
\email{wglevine@epss.ucla.edu}

\author[0000-0003-2215-8485]{Renyu Hu}
\affiliation{Jet Propulsion Laboratory, California Institute of Technology, 4800 Oak Grove Drive, Pasadena, CA 91109, USA}
\affiliation{Division of Geological and Planetary Sciences, California Institute of Technology, Pasadena, CA 91125, USA}
\email{renyu.hu@jpl.nasa.gov}

\author[0000-0002-8958-0683]{Fei Dai}
\affiliation{Institute for Astronomy, University of Hawai'i, 2680 Woodlawn Drive, Honolulu, HI 96822, USA}
\email{fdai@hawaii.edu}

\author[0000-0002-6939-9211]{Tansu~Daylan}
\affiliation{Department of Physics and McDonnell Center for the Space Sciences, Washington University, St. Louis, MO 63130, USA}
\email{tansu@wustl.edu}

\author[0000-0003-2996-8421]{John~P.~Doty}
\affiliation{Noqsi Aerospace Ltd., 15 Blanchard Avenue, Billerica, MA 01821, USA}
\email{jpd@noqsi.com}

\author[0000-0003-1286-5231]{David~R.~Rodriguez}
\affiliation{Space Telescope Science Institute, 3700 San Martin Drive, Baltimore, MD, 21218, USA}
\email{drodriguez@stsci.edu}

\author[0000-0002-6778-7552]{Joseph~D.~Twicken}
\affiliation{SETI Institute, Mountain View, CA 94043 USA/NASA Ames Research Center, Moffett Field, CA 94035 USA}
\email{joseph.twicken@nasa.gov}

\author[0000-0001-9911-7388]{David~W.~Latham}
\affiliation{Center for Astrophysics | Harvard \& Smithsonian, 60 Garden St, Cambridge, MA 02138, USA}
\email{dlatham@cfa.harvard.edu}

\author[0000-0002-4715-9460]{Jon~M.~Jenkins}
\affiliation{NASA Ames Research Center, Moffett Field, CA 94035, USA}
\email{jon.jenkins@nasa.gov}

\author[0000-0001-8227-1020]{Richard~P.~Schwarz}
\affiliation{Center for Astrophysics | Harvard \& Smithsonian, 60 Garden St, Cambridge, MA 02138, USA}
\email{rpschwarz@comcast.net}

\begin{abstract}

We report the validation of a third terrestrial exoplanet in the nearby (22 pc) TOI-2267 system. TOI-2267 is a binary system with stellar components TOI-2267A (M5, 3030 K) and TOI-2267B (M6, 2930 K), with an on-sky separation of 0\farcs384 (8 au projected separation). TOI-2267 hosts two Earth-sized planets (TOI-2267 b,  $1.00\pm0.11 R_{\oplus}$, and TOI-2267 c, $1.14\pm0.13 R_{\oplus}$, if orbiting the primary star; or $1.22\pm0.29 R_{\oplus}$ and $1.36\pm0.33 R_{\oplus}$, respectively, if orbiting the secondary star) with orbital periods of 2.3 and 3.5 days. This system also contains a third Earth-sized planet candidate with an orbital period of 2.0 days that was previously identified as a likely planet with a low false-positive probability, but could not be firmly validated due to the lack of independent observations beyond TESS data. We combine two new transit observations from the 5.1m Hale Telescope at Palomar Observatory with archival TESS data and high-resolution imaging to statistically validate the planetary nature of TOI-2267 d ($0.98\pm0.09 R_{\oplus}$ if orbiting the primary star, or $1.77\pm0.43 R_{\oplus}$ if orbiting the secondary star) using the updated TRICERATOPS+ pipeline. We attempt to determine the host star for TOI-2267 d using transit shape stellar density analysis, but are unable to conclusively assign a host. Our validation of TOI-2267 d suggests that TOI-2267 is either the first known double transiting M dwarf binary system, or hosts three planets in an extremely compact orbital configuration. %around one star (as proposed by Asiru et al. in prep.).

\end{abstract}

%\keywords{\uat{Galaxies}{573} --- \uat{Cosmology}{343} --- \uat{High Energy astrophysics}{739} --- \uat{Interstellar medium}{847} --- \uat{Stellar astronomy}{1583} --- \uat{Solar physics}{1476}}

\section{Introduction} \label{sec:intro}

M dwarfs are the most common type of star in our galaxy. They account for 60–75\% of all stars within 10 pc \citep{Henry2006,Reyle2021}, and $\sim$70\% of all stars in the Milky Way \citep{Bochanski2010}. The relatively small masses, radii, and temperatures of M dwarfs make them uniquely favorable targets for the characterization of terrestrial exoplanets \citep[e.g.,][]{Triaud2021}. Terrestrial planets orbiting M dwarfs have larger planet-to-star mass ratios and corresponding radial velocity semi-amplitudes than terrestrial planets around Sun-like stars. Their larger planet-to-star radius ratios also make it easier to detect their thermal emission and search for signs of atmospheric absorption in transit \citep{Wordsworth2022}. The lower temperatures of M dwarfs also mean that the habitable zone is located at smaller orbital separations, where planets are more likely to transit their host star \citep[e.g.,][]{Nutzman2008,Suissa2020,Gilbert2023}.

Field M dwarfs have a multiplicity rate of 20–30\% \citep[e.g.,][]{Ward-Duong2015,Winters2019,Clark2024}. These stellar companions can alter the formation and evolution of planets in multiple ways. Systems with multiple stars have less massive, shorter-lived, and potentially truncated protoplanetary disks, which can make it harder for planets to accrete gas envelopes \citep[e.g.,][]{Harris2012,Kraus2016,Sullivan2023,Sullivan2024}. The presence of a stellar companion can also dynamically perturb a planet's orbit, potentially resulting in migration or ejection \citep[e.g.,][]{Kaib2013}. For binary systems with small orbital separations, the increased XUV flux may also result in higher mass loss rates, making it more challenging for small planets to retain their atmospheres \citep[e.g.,][]{Johnstone2019,Sullivan2024}. All of these effects should be stronger for close ($< 50$ au) binaries, making these systems especially valuable testbeds for understanding the effects of binarity on planet properties.

A recent survey of M dwarf transiting planet candidates identified by the TESS survey found that $19\pm3\%$ of the planet candidate host stars have stellar companions, consistent with the rate for field M dwarfs \citep{Matson2025}. They also found that the semi-major axis distribution of these stellar companions was shifted to larger values and the mass ratio distribution peaked at lower values as compared to field M dwarfs \citep{Matson2025}.  This suggests that equal-mass binaries with small orbital separations may be less likely to host transiting planets, in good agreement with previous results from surveys of Sun-like stars \citep{Ngo2016,Hirsch2021}. However, to date, there has been relatively little work done on the effect of M dwarf binarity on planet properties. Only two confirmed planet-hosting M dwarfs in binary systems have projected separations $\lesssim$ 50 au (Kepler-289, \cite{Barclay2015}; and K2-288, \cite{Feinstein2019}). Two other similar TESS systems have multiple planet candidates awaiting confirmation (TOI-864 and TOI-3494, \cite{Matson2025}). These rare systems are therefore extremely valuable for understanding how binarity affects planet formation.
%But planets in close binary systems are rare, and can be more challenging to confirm due to the nature of blended light from unresolved companions, so their planet candidates often go without confirmation. 
%There are currently only 6 M dwarf terrestrial planets (within 20\% of 1 $R_{\oplus}$) in multiple star systems \citep{ExoplanetArchive}, and only 2 have mass measurements. Confirmation of additional terrestrial planets in close binary systems is therefore highly valuable, and the TOI-2267 system provides one such opportunity. 

TOI-2267 is a nearby (22 pc) binary star system containing a primary M5, 3030 K component (TOI-2267A) and secondary M6, 2930 K component (TOI-2267B), initially characterized in \cite[][hereafter referred to as ZP25]{Zuniga-Fernandez2025}. TOI-2267A and B have an on-sky separation of 0\farcs384, corresponding to a projected separation of just 8 au. This system comprises two Earth-sized planets with orbital periods of 2.3 and 3.5 days, which are close to the 3:2 mean motion resonance (MMR). ZP25 measured planetary radii of $1.00\pm0.11 R_{\oplus}$ and $1.14\pm0.13 R_{\oplus}$, respectively, if they orbit the primary star, or $1.22\pm0.29 R_{\oplus}$ and $1.36\pm0.33 R_{\oplus}$ if they orbit the secondary star. ZP25 found that the Bayesian evidence from transit fitting does not favor either star, and they were therefore not able to conclusively assign the planets to a host star. The proximity of the two planets to a MMR also strongly suggests that they orbit the same host star, as such a configuration would be very unlikely to occur if each planet orbited a different star with a period randomly drawn from the observed population-level distribution \citep{Hsu2020,Hardegree-Ullman2019,Kaminski2025,Mignon2025}.

The TESS photometry for TOI-2267 also contains a third transit signal with an orbital period of 2.0 days. However, ZP25 were unable to conclusively validate this candidate without ground-based follow-up observations demonstrating that its transit signal occurs on target. Still, statistical validation performed with TRICERATOPS supported the interpretation of the signal as a likely planetary candidate. In this study, we present two new ground-based transits of this candidate observed with Palomar/WIRC and use them to firmly validate the planetary nature of TOI-2267 d. In Section \ref{sec:Observations}, we describe the observations from TESS and Palomar/WIRC. In Section \ref{sec:transit analysis}, we describe our transit analysis and show that the updated false positive probability for this candidate lies below the threshold for statistical validation. In Section \ref{sec:Host star analysis}, we explore whether stellar density profiling can be used to assign TOI-2267 d to a host star. In Section \ref{sec:Discussion}, we summarize our key findings and outline future observations that could further improve our understanding of this system.

\section{Observations} \label{sec:Observations}

\subsection{TESS} \label{sec:TESS Observations}

TOI-2267 was observed in 12 TESS sectors (19, 20, 25, 26, 40, 52, 53, 59, 60, 73, 79, and 86) from December 2019 to December 2024. ZP25 reported the discovery of TOI-2267 b and c at orbital periods of $\sim2.2891$ and $\sim3.4950$ days, respectively, using 2-minute cadence data from the Science Processing Operations Center pipeline \citep[SPOC,][]{Jenkins2016} for all 12 available TESS sectors. A third planet candidate with an orbital period of $\sim$2.0345 days was first identified in the analysis of ZP25 using the \texttt{SHERLOCK}\footnote{\url{https://github.com/franpoz/SHERLOCK}} pipeline \citep{pozuelos2020,devora2024}, based on the initial four TESS sectors available at the time (see ZP25 for further details). This signal had not yet been alerted by SPOC but was later recovered in the SPOC transit search with an adaptive, noise-compensating matched filter \citep{Jenkins2002,Jenkins2010,Jenkins2020} after the ninth TESS sector (sector 60). The TESS Science Office formally released this candidate as TOI-2267.02 on August 15, 2023. In this work,  we refer to it as TOI-2267 d to align with the naming convention of ZP25. We note that this name is not intended to imply that the planet orbits the same star as the other two planets, and we remain agnostic as to which binary component hosts each planet. Once future work confidently assigns stellar hosts, these planets should thereafter be referred to using the standard binary planet nomenclature (e.g., `TOI-2267A d' or `TOI-2267B d').

ZP25 analyzed the TESS target pixel files and SPOC apertures for each sector of TESS observation and verified that very few faint ($\Delta$mag $>$ 5) companion stars were in each aperture, more than 1 TESS pixel away ($>$ 21$\farcs$) from TOI-2267. ZP25 therefore used the Presearch Data Conditioning Simple Aperture Photometry (PDCSAP) flux data, which are corrected for crowding and systematic effects \citep{Stumpe2012,Stumpe2014,Smith2012}, for their subsequent fits. ZP25 performed a global fit including transit light curve models for all three planets and a Gaussian process model for out-of-transit flux variations. They incorporated ground-based observations for TOI-2267 b and c in their joint fit, but were unable to obtain any ground-based transit detections for TOI-2267 d. We refer the reader to ZP25 for further details on this transit modeling. We use the detrended and stacked TESS transit profile from ZP25 which includes 108 total transits of TOI-2267 d in a joint fit with our new ground-based observations (see Section $\ref{sec:transit analysis}$) to statistically validate TOI-2267 d (see Section $\ref{sec:TRICERATOPS analysis}$).

\subsection{Palomar/WIRC} \label{sec:Palomar Observations}

We observed two transits of TOI-2267 d in the $J$-band with the Wide-field Infared Camera (WIRC) on the Hale Telescope at Palomar Observatory, California, USA. The Hale Telescope is a 5.08-m telescope equipped with a 2048 x 2048 Rockwell Hawaii-II NIR detector, providing a field of view of 8\farcm7 × 8\farcm7 with a plate scale of 0.''25 per pixel \citep[WIRC,][]{Wilson2003}. Our data were taken with a beam-shaping diffuser that increased our observing efficiency and improved the photometric precision and guiding stability \citep{Stefansson2017,Vissapragada2020}. 

\begin{figure*}
\begin{center}
  \includegraphics[width=18cm]{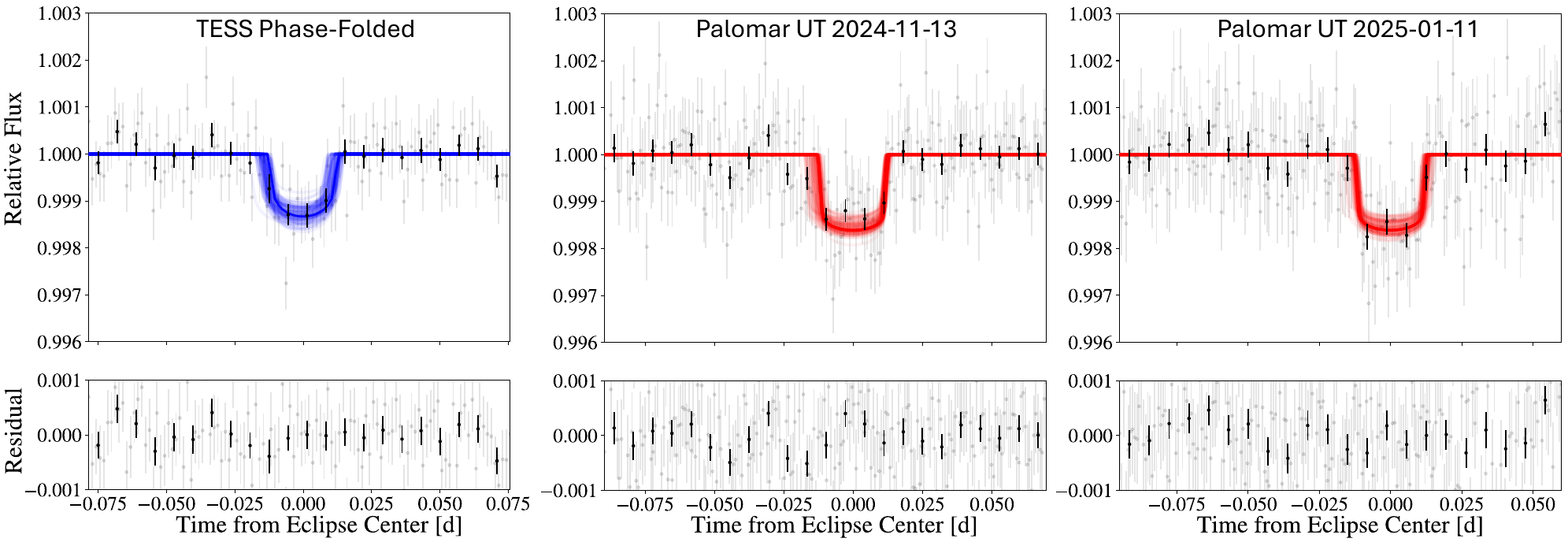}
  \caption{Phased-folded and 2-minute binned TESS light curve (left panel) and detrended Palomar/WIRC light curves (middle and right panels) for the two transit observations of TOI-2267 d. Residuals after the best-fit transit light curve model has been subtracted are shown in the lower panels. Unbinned data are shown as grey circles, with 10 minute binned points overplotted as black circles.  The best joint-fit transit models are overplotted as blue lines for the TESS data and red lines for the Palomar data, with 100 random draws from the posterior distribution to illustrate the typical model uncertainty. All detrended light curves are available electronically as data behind the figure.}
  \label{fig:Transit Plot}
\end{center}
\end{figure*}

We observed transits of TOI-2267 d on UT 2024-11-13 and UT 2025-01-11. We used 4-second exposure times stacked with 9 total co-added exposures per image, and observed full transits plus more than 1 transit duration of baseline both pre-ingress and post-egress on each night (Figure \ref{fig:Transit Plot}). For each night, we obtained calibration images to dark-subtract, flat-field, remove dead and hot pixels, and remove detector structure with a dithered sky background frame following the methodology of \cite{Vissapragada2020}. We extracted photometry and detrended the light curves with the procedure described in \cite{GreklekMcKeon2023}. Conditions were good on both nights, with seeing well below the 3\farcs0 FWHM of our beam-shaping diffuser.  This ensured that PSFs for the target and comparison stars remained stable over the course of our observations.

We began by extracting photometry for our target star and a set of 10 nearby comparison stars. We cleaned the target and comparison light curves by applying a moving median filter with a width of 31 data points corresponding to a time interval of 34 minutes, and removing 5$\sigma$ outliers. We then tested circular photometric apertures with radii ranging between 5 - 25 pixels and selected the optimal aperture radius by minimizing the root mean square scatter after the light-curve fitting described in Section \ref{sec:transit analysis}. Our optimal aperture radii were 20 pixels (5$\farcs$0) and 18 pixels (4$\farcs$0) for UT 2024-11-13 and 2025-01-11, respectively. This meant that both binary components of the TOI-2267 system were fully contained within our photometric aperture.  We provide additional information about these transit observations, including observation times, transit coverage, airmasses, and measured transit mid-times, in Table \ref{tab:Palomar obs}.

\begin{table*}
\centering
\caption{Summary of Palomar/WIRC observations of TOI-2267 d.}
\label{tab:Palomar obs}
\hspace*{-2cm}
\begin{tabular}{cccccccc}
\hline
\hline
UT Date & Start & Finish & Transit \% & Baseline \% & z$_{\text{st}}$ &  z$_{\text{min}}$ & z$_{\text{end}}$
\\ \hline
2024 Nov 13 & 03:53:19 & 07:44:47 & 100\% & 470\% & 1.749 & 1.615 & 1.615
\\ %\hline
2025 Jan 11 & 03:43:52 & 07:27:04 & 100\% & 530\% & 1.617 & 1.610 & 1.654
\\ 
\hline
\end{tabular}
\vspace{0.1cm}
\footnotesize
\begin{flushleft}
\textbf{Notes.}
Start and Finish columns represent the time of the first and last science images in UT time, the transit and baseline fractions are relative to the total transit duration for TOI-2267 d, z$_{\text{min}}$ is the minimum airmass of the science sequence while z$_{\text{st}}$ and z$_{\text{end}}$ are the starting and ending airmasses.
%\tablenotemark{\scriptsize a} Derived from
\end{flushleft}
%Transit Midtime (BJD, \S\ref{sec:transit analysis})
%2460627.7541 $\pm$ 0.0010
%2460686.7538 $\pm$ 0.0011

\end{table*}

\section{Transit Analysis}
\label{sec:transit analysis}

\subsection{Transit Modeling}
\label{subsec: transit modeling}

We used the \texttt{exoplanet} package \citep{exoplanet} to fit each of the two ground-based WIRC light curves with a combined systematics and transit model as described in \cite{GreklekMcKeon2023}. Our systematics model for each night included a linear combination of comparison star light curve weights, an error inflation term added in quadrature to the measured flux errors, and a linear slope. For each night, we extracted photometry for the 10 comparison stars that have minimal variance relative to the time-changing flux of the target star. These initial 10 comparison stars were reduced to 4 for each night because the significance of their weights was more than $1\sigma$ from zero in the final posteriors from our systematics model optimization procedure described below.

We also tested systematics models with linear combinations of weights for the target centroid offset, PSF width, airmass, and local sky background as a function of time. We compared the Bayesian Information Criterion \citep[BIC,][]{Schwarz1978} for all possible combinations of these systematic noise parameters using the same framework as in \cite{Jorge2024}. We found that the model that produced the lowest BIC value included weights for the local sky background and target PSF width on UT 2024-11-13, while our UT 2025-01-11 observations preferred no additional detrending parameters in the systematics model.

When optimizing the systematics model for each night of WIRC data, we also fit for the transit shape parameters (impact parameter $b$, planet-star radius ratio $R_p$/$R_{\star}$, and semi-major axis ratio $a$/$R_*$), along with the stellar radius.  We used wide uniform priors from 0.0 to 1.0 for $b$ and from 0.0 to 0.2 for $R_p$/$R_*$, and normal priors on the stellar radius and $a$/$R_*$ values based on the reported stellar mass, radius, and orbital periods reported in ZP25. We adopted a normal prior on the planetary orbital period $P$ from ZP25, and placed a wide uniform prior on the transit time of $\pm90$ minutes centered on the predicted time using the ephemeris from ZP25. We did not correct the measured $R_p$/$R_{\star}$ parameter for dilution effects from the other binary component in these initial fits, as the purpose of these fits was only to optimize the systematics model for each night and ensure that the results are consistent regardless of stellar host prior choice. 

We performed two fits for each night of data -- one assuming the primary star is the host, and one assuming the secondary star is the host. We adopted stellar parameters from ZP25 for each case and used \texttt{ldtk} \citep{Parviainen_2015} to calculate the $J$-band quadratic WIRC limb darkening parameters, which we held fixed in our fits. In the primary star case, $u_1 = 0.174$ and $u_2 = 0.136$, and in the secondary star case $u_1 = 0.188$ and $u_2 = 0.168$. We explored the parameter space of our model with the NUTS sampler \citep{hoffman2011nouturn} in \texttt{PyMC3} \citep{exoplanet:pymc3} for 2500 tune and 2000 draw steps, and confirmed that the Gelman-Rubin statistic values were < 1.01 for all parameters, indicating good fit convergence.

We confirmed that for both nights of Palomar/WIRC observation, the preferred systematics model is the same regardless of the host star choice, and the $R_p$/$R_{\star}$ constraint is $> 3\sigma$ from 0 indicating a strong detection of the transit around the target. We also confirmed that the measured $R_p$/$R_{\star}$, $a$/$R_{\star}$, and $b$ values are consistent with each other within $1\sigma$ across the two nights and regardless of the stellar host choice.
%and consistent with the reported diluted $R_p$/$R_*$ value from a fit to the TESS data alone in ZP25. 

We conclude that both transits are strongly detected and that our fitted transit shapes are consistent regardless of host star prior choice. We therefore proceeded to jointly fit both Palomar/WIRC transits assuming a common transit shape.  We also confirmed that our fitted $b$ and $a$/$R_*$ values are consistent with the values from the TESS-only fit reported in ZP25.  We therefore incorporated the stacked TESS transit profile into our joint transit fit in order to obtain the most precise constraints on the transit shape parameters. We allowed for separate $R_p$/$R_{\star}$ values in the TESS and WIRC bandpasses in order to determine whether or not these two parameters are consistent after accounting for the wavelength-dependent dilution from the binary. 

We carried out two versions of this joint fit with different assumptions for each. In the first version, we assumed that the primary star is the host, and in the second version we assumed that the secondary star is the host. We used the same model parameters and priors as described in the individual night fits, with $b$ and $a$/$R_*$ now shared across all three transit light curves.  We allowed the individual transit midtimes to vary as free parameters using the \texttt{TTVOrbit} module of \texttt{exoplanet}, with the stacked TESS transit profile arbitrarily shifted to 1 orbital period before the first Palomar transit. We allow the transit midtimes to freely vary due to the potential for TTVs in the system (see ZP25 and Section \ref{sec:Host star prospects}). We also fit for the true (undiluted) $R_p$/$R_{\star}$ value in each bandpass (TESS and $J$). At each step of the fit, we converted the true $R_p$/$R_{\star}$ values to predicted transit depths in each bandpass using a radius correction factor $X_R$ incorporating the wavelength-dependent flux dilution from the companion. The radius correction factor was calculated using Equation 7 of \cite{Ciardi2015} for the primary star case, and Equation 6 of \cite{Ciardi2015} for the secondary star case.  We assumed a flux ratio $F_{\text{secondary}}/F_{\text{primary}}$ of $0.3045\pm0.05$ in the TESS bandpass and $0.3571\pm0.05$ in the $J$ band, as reported in ZP25.  We accounted for the uncertainties on these two flux ratios by making them both free parameters in our fit and placing a Gaussian prior on each that matches the value reported by ZP25.

We explored the parameter space of our joint transit model with the NUTS sampler in \texttt{PyMC3} for $10^4$ tune and $10^4$ draw steps, and confirmed that the Gelman-Rubin statistic values are < 1.01 for all parameters, indicating that the fit has converged. The posterior distributions for our transit model parameters and related derived quantities for both the primary and secondary star fits are summarized in Table \ref{table:planet_params_d}, and corner plots illustrating the posterior distributions for all joint model parameters, including systematics, are included in the appendix.

\begin{table*} 
\centering
\resizebox{0.8\textwidth}{!}{%
\begin{threeparttable}
\begin{tabular}{l c c r}
\toprule
Parameter & Unit & Primary host & Secondary host \\ \midrule
% &  & \multicolumn{2}{c}{TOI-2267\,b}\\
\midrule
\multicolumn{2}{c}{\it{Model Parameters}} & \multicolumn{2}{c}{TOI-2267 d}  \\
\midrule \smallskip 
Undiluted $R_p / R_{\star}$ TESS &  &  $0.041\pm0.003$ & $0.121_{-0.030}^{+0.032}$ \\ \smallskip
Radius correction factor \textit{TESS}, $X_\mathrm{R;\,\textit{TESS}}$ & & $1.142\pm0.021$ & $3.547_{-0.867}^{+0.861}$ \\ \smallskip
Undiluted $R_p / R_{\star}$ WIRC &  &  $0.046\pm0.002$ & $0.129_{-0.031}^{+0.032}$ \\ \smallskip
Radius correction factor \textit{WIRC}, $X_\mathrm{R;\,\textit{WIRC}}$ & & $1.166\pm0.021$ & $3.349_{-0.810}^{+0.808}$ \\ \smallskip
$a/R_*$ & & $18.617_{-2.570}^{+2.625}$ & $22.693_{-5.967}^{+3.024}$ \\ \smallskip
Impact Parameter $b$ & & $0.697_{-0.128}^{+0.084}$ & $0.449_{-0.298}^{+0.302}$ \\ \smallskip
%Mid-transit time, $T_0$ TESS & BJD$_{TDB}$-2457000 & $1817.0840_{-0.0013}^{+0.0017}$ & $1817.0840_{-0.0014}^{+0.0016}$ \\  \smallskip
Mid-transit time, $T_0$ WIRC Night 1 & BJD$_{TDB}$-2457000 & $3627.7542_{-0.0010}^{+0.0007}$ & $3627.7544_{-0.0009}^{+0.0006}$ \\  \smallskip
Mid-transit time, $T_0$ WIRC Night 2 & BJD$_{TDB}$-2457000 & $3686.7539\pm0.0007$ & $3686.7539_{-0.0007}^{+0.0006}$ \\
\midrule
\multicolumn{2}{c}{\it{Derived Parameters}} & & \\
\midrule \smallskip
$^{a}$Planet Radius, $R_p$ & $R_\oplus$ & $0.98\pm0.09$ & $1.77\pm0.43$ \\ \smallskip
Semimajor axis, $a$ & au & $0.0174\pm0.0003$ & $0.0145_{-0.0007}^{+0.0006}$ \\ \smallskip
Orbital period,  $P$ & days &  $2.0344678\pm2.3\times10^{-6}$ & $2.0344671\pm2.3\times10^{-6}$ \\ \smallskip
Mid-transit time, $T_0$ & BJD$_{TDB}$-2457000 & $3625.7199_{-0.0015}^{+0.0012}$ & $3625.7200_{-0.0017}^{+0.0013}$ \\ \smallskip
Inclination, $i$ & $^{\circ}$ &  $87.4\pm0.6$  & $89.0_{-0.7}^{+0.6}$ \\ \smallskip
%Transit duration, $T_{1-4}$ & hrs &  $0.671_{-0.059}^{+0.049}$ & $0.709_{-0.060}^{+0.051}$ \\ \smallskip
$^{b}$Equilibrium Temperature, $T_{eq}$ & K & $503\pm32$ & $422_{-56}^{+53}$ \\ \smallskip
Insolation Flux, $S$ & S$_{\oplus}$ & $10.7_{-2.5}^{+3.0}$ & $5.3_{-2.3}^{+3.2}$  \\ \smallskip
Transmission spectroscopy metric (TSM)$^{a}$ & & $18.9_{-5.8}^{+8.1}$ & $267_{-163}^{+352}$  \\ \smallskip
%Emission spectroscopy metric (ESM) & & $8.6_{-3.3}^{+6.4}$ & $7.6_{-4.1}^{+10.5}$  \\ \smallskip
Host density from transit, $\rho_\mathrm{\star}$ & g\,cm$^{-3}$ & $32.0_{-10.8}^{+14.7}$ & $60.4_{-35.7}^{+26.3}$ \\
\midrule 
\end{tabular}
\end{threeparttable}
}
\caption{Model and derived parameters for the joint TESS+Palomar transit fitting of TOI-2267 d, for cases where the primary or the secondary star is the host. Adapted from ZP25. \\
$^{a}$Calculated from the stellar radius values reported in ZP95 ($R_{\text{primary}}=0.2075\pm0.0225$, $R_{\text{secondary}}=0.130\pm0.030$) and taking the error-weighted average across the two bands. \\
$^{b}$Values calculated assuming an albedo of 0.3 (Earth-like) and an Earth-like bulk density, from \cite{Kempton2018}. }
\label{table:planet_params_d}
\end{table*}

\subsection{Planet Validation} \label{sec:TRICERATOPS analysis}

We find that for both the primary and the secondary star cases, the dilution-corrected planet-star radius ratio for TOI-2267 d is consistent at the $1\sigma$ level between the TESS and $J$ bands. This increases our confidence that this candidate is a genuine planet, as stellar eclipsing binaries often exhibit wavelength-dependent transit depths. In this section, we quantify the effect of this knowledge on the planet's False Positive Probability (FPP) and the Nearby False Positive Probability (NFPP) using the \texttt{TRICERATOPS+} statistical validation code. The \texttt{TRICERATOPS+} FPP describes the probability that a transit signal does not originate from a planet transiting the target star. The NFPP describes the probability that the observed transit signal originates from a resolved nearby star (i.e., a star farther away than 1$\farcs$2 that contributes sufficient flux to the TESS aperture to produce the observed signal) rather than the target star. The FPP encapsulates all possible false positive scenarios, whereas the NFPP considers only a subset of them. %Thus, the NFPP will always be less than or equal to the FPP. 

The original \texttt{TRICERATOPS} code \citep{Giacalone2021} models the light curves of transiting planets and various astrophysical false positive scenarios, incorporating prior information about the population of stars in the Milky Way to compute the FPP and NFPP for a given planet candidate. This code has been widely used to confirm TESS planet candidates \citep[e.g.,][]{Thomas2025,Scott2025,Barkaoui2025,Stalport2025}. In \texttt{TRICERATOPS+} \citep{GomezBarrientos2025}\footnote{\url{https://github.com/JGB276/TRICERATOPS-plus}}, we updated this code to incorporate information from transit light curves obtained in multiple bandpasses. 

\texttt{TRICERATOPS+} does not directly utilize any of the information from our transit light curve fits in \S\ref{subsec: transit modeling}.  Instead, it generates a series of scenarios for both transiting planet and stellar eclipsing binary cases by drawing parameters randomly from a set of prior distributions informed by observations of both populations.  For each randomly generated scenario, it calculates a predicted bandpass-specific light curve.  It then compares this predicted light curve to the measured light curve in that bandpass and calculates the corresponding log-likelihood value (Equation 16 of \citealt{Giacalone2021}).  The code then sums the individual log-likelihoods for each bandpass to make a single combined log-likelihood for that scenario.  The resulting ensemble of log-likelihoods is then used to calculate the FPP and NFPP as described in \cite{Giacalone2021}.  This package calculates limb darkening coefficients using the \texttt{ExoTIC-LD} package \citep{Grant2022}; in order to do so for TOI-2267, we utilize the stellar parameter values reported in Table 2 of ZP25. For scenarios involving eclipsing binaries and/or contamination from unresolved companions, as is the case for the TOI-2267 system, \texttt{TRICERATOPS+} also incorporates the bandpass-specific flux ratio when computing predicted model light curves for each scenario \citep[for more details, see][]{GomezBarrientos2025}. %We fix the TESS and $J$ band flux ratios to the best-fit values reported in ZP25 for TOI-2267. 

We computed the FPP and NFPP for TOI-2267 d using the phased TESS transit light curve, both Palomar light curves, and high-resolution contrast curves from the ‘Alopeke speckle instrument on the Gemini North 8-m telescope and the speckle polarimeter on the 2.5-m telescope at the Caucasian Observa- tory of Sternberg Astronomical Institute, both of which were published in ZP25. This observation detected the companion star TOI-2267B at a separation of 0$\farcs$384, and found no evidence for any additional companions brighter than 4-5 magnitudes below that of TOI-2267A out to a distance of 1$\farcs$2. 

In the \texttt{TRICERATOPS} analysis of ZP25, multiple terms are excluded from the NFPP calculation, which correspond to scenarios that have already been ruled out for the TOI-2267 system. These include scenarios where the star does not have an unresolved companion, and scenarios where there is a background star at the current location of TOI-2267, which was ruled out through an analysis of archival images for this high proper motion star. Removing these terms has the effect of decreasing the NFPP value \citep[see equation 5 of][]{Giacalone2021}. ZP25 also added two new terms to the FPP calculation, corresponding to scenarios with a transiting planet around an unresolved bound companion \citep[see equation 4 of][]{Giacalone2021}, which has the effect of decreasing the FPP value. We do not modify the FPP and NFPP calculations, instead adopting the standard approach described in \cite{Giacalone2021}.  This provides us with a conservative upper limit on the false positive probabilities for TOI-2267 d. Following previous statistical validation studies with \texttt{TRICERATOPS} \citep[e.g.,][]{Giacalone2021, Giacalone2022}, we ran the analysis 20 times and report the average value and the 68$\%$ confidence interval of the FPP and the NFPP. We obtain a FPP of $4.7\times10^{-6}$, with a 68\% confidence interval range of ($4.5\times10^{-7}, 5.0\times10^{-5}$). We calculate a NFPP of 0; this reflects the fact that our Palomar observations detected transits with the expected depth around the target system, which is the unresolved light of both TOI-2267A and B. These detections indicate that the signal is coming from either TOI-2267A or B, allowing us to confidently rule out events that occur on other nearby spatially resolved stars whose light is blended with the target system in the TESS aperture.  We do not consider the scenario where the planet transits the secondary star to be a “nearby false positive” in the traditional sense, because this analysis still supports the hypothesis that TOI-2267 d must be a planet orbiting either TOI-2267A or B. As defined in \citealt{Giacalone2021}, the recommended threshold to validate a planetary candidate is FPP $< 1.5\%$ and NFPP $< 0.1\%$. This means that TOI-2267 d definitively satisfies the requirements for a statistically validated planet orbiting one of the two stars.

%\textbf{We also performed a second version of the NFPP calculation, with a modified definition from the original \texttt{TRICERATOPS} pipeline, which typically only includes stars within the TESS aperture that are resolved by Gaia in the NFPP calculation. For this modified NFPP calculation, we add TOI-2267B to the list of resolved stars that are considered in the analysis. In this case, we get an NFPP of 0.14, with a 68\% confidence interval range of (0.13, 0.15). This value is entirely driven by the scenario where there is a transiting planet around TOI-2267B. We caution that we do not consider this a “nearby false positive” in the traditional sense because this analysis still supports the hypothesis that TOI-2267 d is orbiting either TOI-2267A or B, but it demonstrates that the photometric data alone are not enough to conclusively confirm or deny the secondary component as the host star for TOI-2267 d.} 

\begin{figure}
\begin{center}
  \includegraphics[width=8cm]{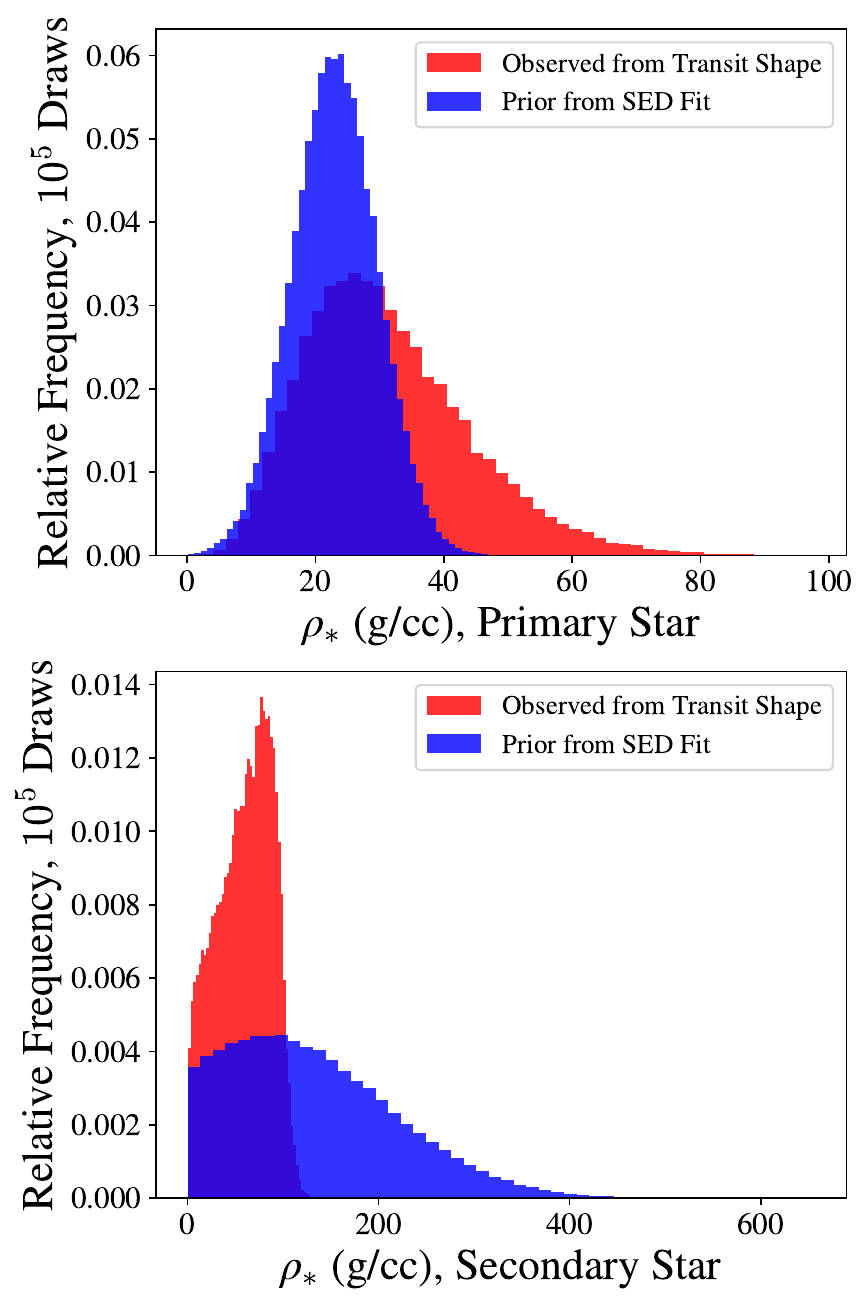}
  \caption{Empirical stellar density constraints derived from our transit shape analysis (red) compared to the stellar density measurements from the SED fitting of ZP25 (blue), for the primary (top) and secondary star (bottom) fits. Our empirical constraints on the stellar density are not precise enough to conclusively rule out either star as the host of TOI-2267 d.}
  \label{fig:Stellar density plot}
\end{center}
\end{figure}

\section{Host Star Identification} \label{sec:Host star analysis}

%\subsection{Transit Shape Analysis}

Having confirmed the planetary nature of TOI-2267~d, we next attempted to determine which binary component it orbits. First, we compared the quality of the joint transit fit from \S\ref{subsec: transit modeling} for the primary versus secondary star case. Since both cases have the same model framework and number of parameters, we compared the ratio of maximum a posteriori likelihood values identified from each \texttt{PyMC3} sampler chain. We measured a likelihood ratio of 1.5 in favor of the primary star over the secondary star case. However, this preference is not statistically significant; a likelihood ratio closer to 10 is typically required in order to provide strong evidence for one model over another \citep{Deeks168}.

We also attempted to identify the host star using constraints on the stellar density derived from our measurement of the transit shape \citep[e.g.,][]{Lester2022}. We combined our retrieved posterior distribution for $a/R_*$ with the orbital period and the measured stellar radius reported by ZP25 to calculate a posterior probability distribution for the mean stellar density. We repeated this process for fits where the planet was assumed to orbit the primary and where it orbited the secondary. We then compared the inferred stellar densities from each transit fit to the independently measured stellar densities from the SED fitting of ZP25. In previous studies using this technique, the star whose measured stellar density from the transit light curve more closely matched expectations from stellar population models was assumed to be the planet host \citep{Lester2022}. Unfortunately, the density of TOI-2267B is only loosely constrained by SED fitting ($\rho_{*_A} = 23.0^{+5.3}_{-4.0}$, $\rho_{*_B} = 83^{+113}_{-39}$, ZP 25). We found that our empirically constrained stellar densities are consistent with the SED priors in both cases (see Fig. \ref{fig:Stellar density plot}).  We conclude that our combined transit light curves are not precise enough to identify the host star for TOI-2267 d. 

In principle, we can also use the measured occurrence rate as a function of planet radius for low-mass stars as indirect evidence in favor of one binary host over the other. For low-mass M dwarfs such as TOI-2267A and B, TESS results have shown that sub-Neptune sized planets (1.5-2 $R_{\oplus}$) are relatively rare (occurrence rates $\leq 7.3\%$).  In these systems, terrestrial planets ($<$ 1.5 $R_{\oplus}$) outnumber sub-Neptunes by a ratio of 14 to 1 \citep{Ment2023}. Our analysis indicates that TOI-2267 has a radius of 1.77 $\pm 0.43 R_{\oplus}$ if it orbits the secondary star and 0.98 $\pm 0.09 R_{\oplus}$ if it orbits the primary, suggesting that the secondary star scenario may be less likely from a population statistics perspective. Unfortunately, even this qualitative analysis is not conclusive because TOI-2267 d's radius is $1\sigma$ consistent with the terrestrial planet threshold for the secondary host star scenario.
%Additional transit observations of TOI-2267 d or updated stellar characterization of TOI-2267B that improves this radius precision would be able to probabilistically distinguish the host star more conclusively.}

\section{Discussion} \label{sec:Discussion}

\subsection{Prospects for Identifying a Host Star}
\label{sec:Host star prospects}

In the future, improved stellar density constraints might make it possible to unambiguously assign TOI-2267 d to a host star.  However, there are several alternative approaches that could also provide a definitive identification. The most straightforward solution would be to obtain a transit observation of TOI-2267 d that spatially resolves the two stars; however, ground-based AO imaging has not demonstrated that it can achieve the required photometric precision \citep[e.g.,][]{Howell2019}, and this would therefore likely require space-based observations with HST or JWST.  ZP25 also pointed out that significantly higher precision measurements of the wavelength-dependent transit shape after correcting for dilution effects could also be used to assign a stellar host.

In principle, dynamical stability arguments can also be used to eliminate some possible orbital configurations, as the three planets have very similar orbital periods. ZP25 used the SPOCK stability classifier \citep{SPOCK}, a machine-learning model trained on numerical integrations of three-planet systems, to predict the stability of the system if all three planets orbit either the primary or secondary star. They found that these kinds of architectures would be highly unstable, and suggested that all three planets are therefore unlikely to orbit the same star. ZP25 then carried out stability analyses for two-planet scenarios using \texttt{rebound} \citep{rebound} n-body integrations that sample orbital parameters from the posterior distributions reported in ZP25, integrate over $10^7$ orbital periods of the outermost planet, and calculate the MEGNO parameter \citep[Mean Exponential Growth factor of Nearby Orbits;][]{Cincotta1999,Cincotta2003}, a well-established diagnostic for
assessing the dynamical stability of planetary systems \citep[e.g.,][]{Hinse2015,Jenkins2019,Delrez2021}. This analysis found highly stable configurations when TOI-2267 d and c or TOI-2267 b and c orbit the same star, and highly unstable configurations when TOI-2267 d and b orbit the same star, further strengthening the argument that all three planets cannot share one stellar host. ZP25 additionally found that in all of their dynamical simulations of the b-c pair, the system exhibited apsidal libration around a fixed point, suggesting that planets b and c may be locked in the 3:2 mean motion resonance.
However, ZP25 did not test whether the b-d pair might also be locked in an orbital resonance, which could allow for a stable three-planet configuration around one star despite the extremely close orbits. This scenario was originally proposed by Asiru et al. (in prep), who explore the viability of this configuration in more detail using n-body simulations.
%Since attempts to assign a stellar host based on transit data in this work and in ZP25 were inconclusive, and future work in Asiru et al. (in prep.) will investigate an alternative scenario to the non-resonant dynamical arguments for different stellar hosts of the b-d pair presented in ZP25, we remain agnostic as to whether TOI-2267 d shares a host star with TOI-2267 b and/or c.} 

As a result of their compact orbits, any planet pairs orbiting the same star should exhibit transit timing variations. ZP25 predicted that the TTV amplitudes of TOI-2267 b and c due to proximity to the 3:2 resonance are likely to be small ($\lesssim$ 4 min if orbiting the primary, $\lesssim$ 10 min if orbiting the secondary), and are therefore unlikely to be detected using the individual TESS transits times, which have a relatively low SNR. For TOI-2267 d, our individual Palomar/WIRC transit observations achieve a much higher timing precision (1.3~min) than for individual TESS transit observations (median precision of $\sim$12 min). In the future, additional ground-based transit timing measurements of all three planets with a comparable precision could be used to carry out a much more sensitive search for transit timing variations. If all three planets orbit the same star in a resonant configuration, then the TTVs should be significantly larger, allowing for a straightforward test of this hypothesis.
%Asiru et al. (in prep.) originally posed this scenario and will investigate it in detail. No matter what the true dynamical configuration of the TOI-2267 system is, the timing precision from our Palomar/WIRC observations is high enough to improve the TTV models and therefore help reveal the true system architecture once combined with additional transit observations.} 
If TTVs are detected for any of the planets in the TOI-2267 system, the information on planet-to-star mass ratio could also be combined with information on planet-to-star radius ratio and compatibility with each binary component's density to obtain improved constraints on which star the planets orbit.

\subsection{Could TOI-2267 d Host a Volatile-Rich Envelope?} 

Small M dwarf planets may be more susceptible to atmospheric mass loss. M dwarfs have higher fractional XUV fluxes, more frequent flares than their sun-like counterparts, and a longer activity lifetime, which could result in enhanced atmospheric mass loss rates \citep{Johnstone2020,Harbach2021,Atri2021}. The observed flaring in TESS photometry and very short rotation periods of the TOI-2267 stars ($\sim$ 17 hours for TOI-2267A, and $\sim$15 hours for TOI-2267B, ZP25), suggest both stars may have strong stellar winds and energetic particle emissions. Whether TOI-2267 d orbits the primary or the secondary star, its short orbital period indicates that it is likely subjected to extreme space weather conditions. If TOI-2267 d orbits the primary star it would have a radius of $0.98 \pm 0.09$ R$_{\oplus}$ and a predicted equilibrium temperature of $503\pm32$~K, incorporating only the flux from the primary star since the true separation of the binary components is unknown. This makes it a close analogue of Trappist-1~b, which was recently shown to have lost most or all of its atmosphere \citep{Greene2023}.  
%with a tidally locked bright day side that is accessible to surface characterization though emission spectroscopy \citep[e.g.,][]{Paragas2025}.

If TOI-2267~d instead orbits the secondary star, it would have a radius of $1.77 \pm 0.43$ R$_{\oplus}$ and an equilibrium temperature of $422\pm55$~K. Population-level studies of M dwarf planet masses indicate that planets of this size often host volatile-rich envelopes \citep[e.g.,][]{Luque_2022,Rogers_2023}. Regardless of which star TOI-2267 d orbits, with a projected separation of just 8 au the companion star may be close enough to contribute to the planet's stellar radiation environment, with corresponding implications for its atmospheric mass loss history \citep[e.g.,][]{Lammer2013,Cherenkov2017,Alvarado-Gomez2022}. We conclude that this system presents an exciting opportunity for future studies on the impact of stellar multiplicity on space weather and atmospheric mass loss for sub-Neptune-sized exoplanets, if the host star for TOI-2267~d can be identified.

\section{Conclusions} \label{sec:Conclusions}

By combining ground-based observations with statistical validation, we firmly confirm the planetary nature of TOI-2267 d, the third Earth-sized exoplanet identified in the TOI-2267 binary system. However, we are unable to unambiguously determine its host star. If it orbits TOI-2267A, it is a $\sim1.0$ R$_{\oplus}$ terrestrial planet with a predicted equilibrium temperature of approximately 500~K.  Given the relatively high activity levels of both stars and their small projected separation, this would suggest that it has likely lost any primordial atmosphere. If it instead orbits the secondary, it is a $\sim1.8$ R$_{\oplus}$ planet with an equilibrium temperature of approximately 420~K, and therefore might host a volatile-rich envelope. If present, any volatile-rich envelope is likely to experience ongoing atmospheric mass loss driven by the irradiation and winds from both stars. 

Regardless of whether TOI-2267 d orbits the primary or secondary star, the presence of three transiting planets in this system provides an exciting opportunity to explore the effect of a close binary companion on the formation and evolution of sub-Neptune-sized planets. If TOI-2267 d orbits a different star than TOI-2267 b and c, then this system is the first double-transiting binary M dwarf system and the most compact binary system known to host planets around both stars. If all three planets instead orbit one star in a resonant configuration as proposed by Asiru et al. (in prep.), then TOI-2267 would host the most compact orbital configuration of any exoplanet system and have an extremely dynamically delicate architecture similar to that of TRAPPIST-1 \citep{Luger2017,Tamayo2017,Papaloizou2018}. Any detection of transit timing variations (TTVs) for TOI-2267 b, c, or d would prove that they are gravitationally interacting and therefore share a host star. Planet-to-star mass ratios from TTVs could also be combined with planet-to-star radius ratios from the stacked transit profiles to perform updated stellar density analysis and assign a host star \citep[e.g.,][]{Lester2022}. There is therefore great value in obtaining additional high-precision ground-based transit observations of all three planets in this exciting new system.

\section{Acknowledgments}

We thank the Palomar Observatory telescope operators, support astronomers, hospitality, and administrative staff, without whom this research would not have been possible. We are especially grateful to Paul Nied, Carolyn Heffner, Kathleen Koviak, Diana Roderick, and Rigel Rafto who supported our observations of TOI-2267, and to Monastery keeper Jeff. Part of this program was supported by JPL Hale telescope time allocations. We are thankful to the PARVI team and Palomar Observatory directorate, especially David Ciardi, Chas Beichman, Aurora Kesseli, and Andy Boden for their gracious support of the Palomar TTV survey program during periods requiring quick readjustment of the 200-inch observing schedule.

This research was supported from the Wilf Family Discovery Fund in Space and Planetary Science established by the Wilf Family Foundation. F.J.P acknowledges financial support from the Severo Ochoa grant CEX2021-001131-S MICIU/AEI/10.13039/501100011033 and 
Ministerio de Ciencia e Innovación through the project PID2022-137241NB-C43. T.D. acknowledges support from the McDonnell Center for the Space Sciences at Washington University in St. Louis. This research has made use of the NASA Exoplanet Archive \citep{ps} and the Exoplanet Follow-up Observation Program website, which are operated by the California Institute of Technology, under contract with the National Aeronautics and Space Administration under the Exoplanet Exploration Program. The research made use of the Swarthmore transit finder online tool \citep{SwarthmoreTTF}. This paper includes data collected by the TESS mission that are publicly available from the Mikulski Archive for Space Telescopes (MAST). We acknowledge the use of public TESS data from pipelines at the TESS Science Office and at the TESS Science Processing Operations Center. Funding for the TESS mission is provided by NASA's Science Mission Directorate.

\software{\texttt{exoplanet} \citep{exoplanet:joss,
exoplanet:zenodo} and its dependencies \citep{exoplanet:foremanmackey17,
exoplanet:foremanmackey18, exoplanet:agol20, exoplanet:arviz,
exoplanet:astropy13, exoplanet:astropy18, exoplanet:luger18, exoplanet:pymc3,
exoplanet:theano}
\texttt{astropy} \citep{astropy},
\texttt{scipy} \citep{scipy},
\texttt{numpy} \citep{numpy},
\texttt{matplotlib} \citep{matplotlib},
\texttt{rebound} \citep{Tamayo2020}, 
\texttt{reboundx} \citep{Lu_2023},
\texttt{BATMAN} \citep{batman},
\texttt{emcee} \citep{emcee},
\texttt{corner} \citep{corner},
\texttt{lightkurve} \citep{lightkurve}, and
\texttt{Claude 3.5 Sonnet.}}

\facilities{ADS, NASA Exoplanet Archive, ExoFOP, TESS, Hale (Palomar 200-inch/WIRC)}


%\vspace{5mm}




\appendix \label{Appendix}

\begin{figure*}
\begin{center}
  \includegraphics[width=18.0cm]{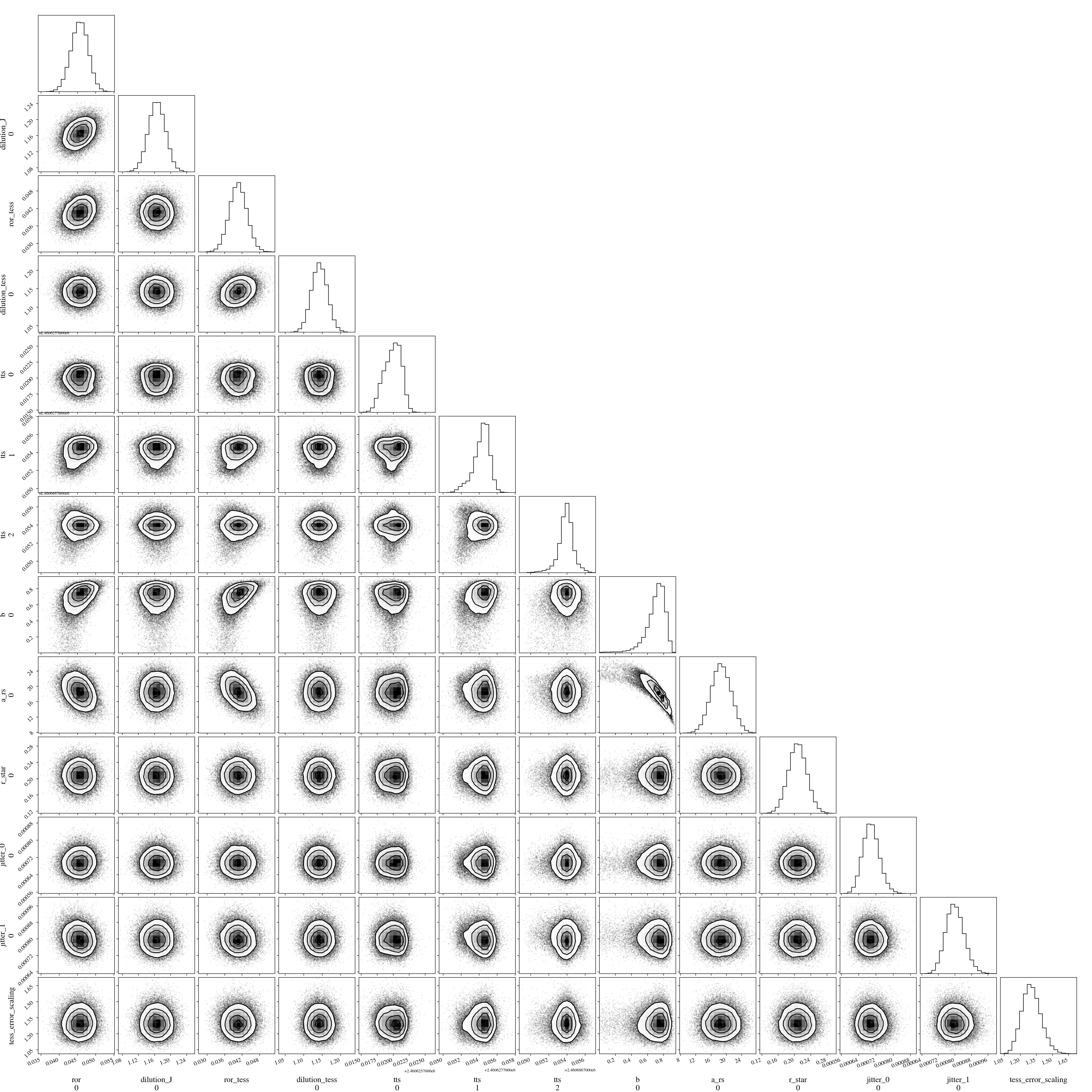}
  \caption{Posterior distribution of transit model parameters in our joint transit fit, assuming the primary host star, using both nights of Palomar/WIRC data and the phase-folded TESS photometry for TOI-2267 d.}
  \label{fig:FullCorner_Primary}
\end{center}
\end{figure*}

\begin{figure*}
\begin{center}
  \includegraphics[width=18.0cm]{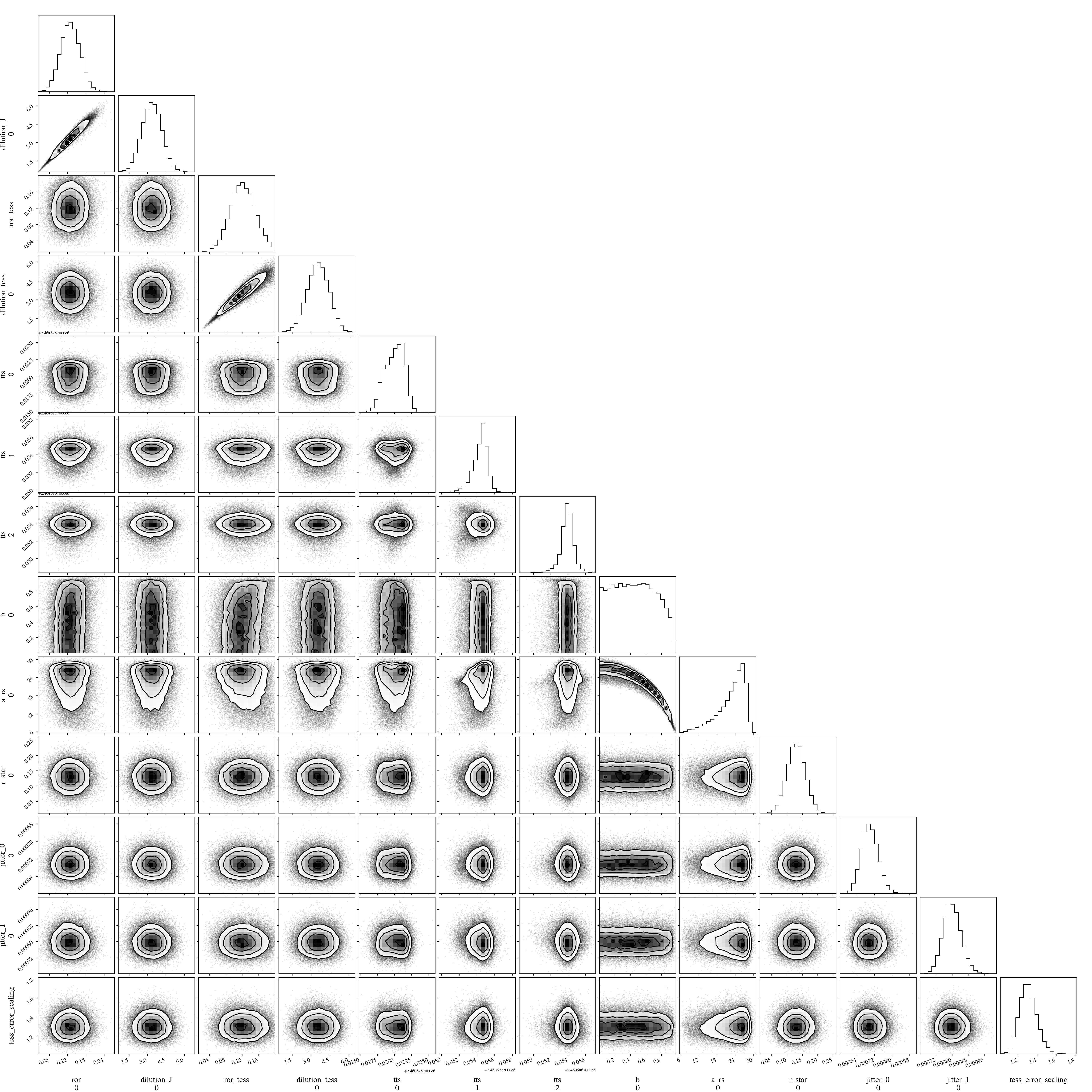}
  \caption{Posterior distribution of transit model parameters in our joint transit fit, assuming the secondary host star, using both nights of Palomar/WIRC data and the phase-folded TESS photometry for TOI-2267 d.}
  \label{fig:FullCorner_Secondary}
\end{center}
\end{figure*}

We present the full posterior distribution for all transit model parameters in the primary star joint fit in Figure \ref{fig:FullCorner_Primary}, and the posterior distribution for all transit model parameters in the secondary star joint fit in Figure \ref{fig:FullCorner_Secondary}.

\bibliography{every_citation_ever}{}
\bibliographystyle{aasjournal}

\end{document}